\magnification=\magstep1
\baselineskip = 12pt
\parskip= 2 pt plus 1pt minus 1pt
\def\emph#1{{\it #1\/}}
\font\title = cmbx10 scaled 1440        % For big title
\newcount\ftnumber
\def\ft#1{\global\advance\ftnumber by 1
          {\baselineskip=13pt \footnote{$^{\the\ftnumber}$}{#1 }}}

\centerline{\title{Annotated Interview with a QBist in the Making}}
\vskip 15pt
\centerline{N. David Mermin} 
\vskip 10pt
\centerline{Laboratory of Atomic and Solid State Physics}
\centerline{Cornell University}
\centerline{Ithaca, NY 14853-2501}
\vskip 5pt
\centerline{and}
\vskip 5pt
\centerline{Stellenbosch Institute for Advanced Study (STIAS)}
\centerline{Jonkershoekweg 19}
\centerline{Stellenbosch 7600}
\centerline{South Africa}
\vskip 20pt

{

\narrower \narrower

\noindent These are my published answers to the seventeen questions posed by Max
Schlosshauer in his 2011 book {\it Elegance and Enigma: the Quantum Interviews\/}.  I 
have inserted thirty footnotes into those answers, commenting on them from the perspective of
subsequent insights I acquired during six weeks in March and April of 2012 with Chris Fuchs  and Ruediger Schack at the Stellenbosch Institute for Advanced Study.   Aside from the footnotes and a few introductory
paragraphs, the text is identical to my contributions to the Schlosshauer volume.

}
\bigskip

My title imitates that of arXiv:1207.2141, ``Interview with a Quantum
Bayesian'', by Christ\-opher A. Fuchs.  Mine differs from his in several 
ways.  I prefer Fuchs's term ``QBist'' because Fuchs's view of
quantum mechanics differs from others as radically as cubism differs
from renaissance painting, and because I find his term ``quantum
Bayesianism'' too broad.  QBism explores the consequences for the
interpretation of quantum mechanics of a thorough-going subjective
view of probability, as an agent's measure of her own personal degree of 
belief.  But there are Bayesians who take probabilities to reflect objective
facts about an event rather than subjective judgments of an agent.  I
would prefer to name Fuchs's perspective not after Thomas Bayes, but
after that eloquent pioneer of subjective probability, Bruno de
Finetti: Quantum Bruno-de-Finetti-ism.  Still QBism, but with the ``B''
for Bruno, not Bayes.

Another difference in our titles is that while Fuchs describes himself
as a QBist, the author of the text below was not yet there.  Max
Schlosshauer asked seventeen questions of seventeen people
and published the 289 answers, question by question, in
his book {\it Elegance and Enigma: The Quantum Interviews\/},
Springer, 2011.  My answers to Schlosshauer's questions were finished in April
2011, a year before I spent six weeks in Stellenbosch at STIAS with
Fuchs and Ruediger Schack, and finally began to understand what they
had been trying to tell me for the past ten years.  (I published 
a summary of my Stellenbosch epiphany as a Commentary
in the July 2012 Physics Today.  That Commentary elicited
several letters to the editor, mostly critical, which were published 
in December 2012, together with my further comments.)

In July 2012, Fuchs posted his own answers to Schlosshauer's 
seventeen questions at arXiv and urged me to post mine.  I hesitated.
By posting nothing at arXiv since 2008, I had ceased to qualify as an
endorser, an appropriate loss of status for a retired country gentleman
that I was anxious not to reverse.  Recently the authorities at arXiv
removed this concern by accepting my request to make my dequalification
permanent. I was also worried that many of the views I expressed in
Schlosshauer's book had changed since Stellenbosch.  But, rather to my
surprise, I discovered that few of them had.  Fortunately
Schlosshauer prohibited footnotes in his book, making it easy for me
to maintain verbatim my original 2011 text, while comparing it 
with my post-Stellenbosch views of early 2013. 

My annotated replies to Schlosshauer's questions are as follows:  
\vskip 25pt

{\centerline{\title The Seventeen Questions of Maximilian Schlosshauer}} 

\bigskip 

\vskip 5pt \noindent{\bf Q1. What first stimulated your interest in the foundations of quantum
mechanics?}

 I've always been more fascinated by physics as a conceptual structure
 than by physics as a set of rules for calculating the behavior of the
 natural world---what Suman Seth calls the ``physics of principles,''
 as opposed to the ``physics of problems.'' My text with Neil Ashcroft
 on solid-state physics is a success because Neil is as focused on the
 physics of problems as I am on the physics of principles. Somehow we
 managed to produce a book that combines both views.

The conceptual structure of quantum mechanics is stranger and lovelier
than any perspective on the world that I know of, so I've been
fascinated and worried about it from the beginning of my career in
physics. Indeed I became interested in my early teens in the late
1940s long before I knew enough mathematics to learn the quantum
formalism, through the popular writings of George Gamow, Arthur
Eddington, and James Jeans. In college, I put these interests on hold,
majoring in mathematics and taking only a few courses in (classical)
physics.

But I returned to physics in graduate school, where my revived
curiosity about quantum foundations was actively discouraged by my
teachers. To my disappointment, the Harvard physicists all believed
that a preliminary training in the physics of problems was a
prerequisite to any understanding of the physics of principles. So,
for a quarter of a century, I was deflected full-time into statistical
physics, low-temperature physics, and solid-state physics, using (and
teaching) quantum mechanics as a beautiful and effective body of rules
for manipulating symbols on a page to get answers to questions about
experiments in the laboratory. ``Shut up and calculate!''

Early in graduate school, Gordon Baym, a fellow student, told me at
the Hayes--Bickford cafeteria about Bohm's spin-1/2 version
of Einstein, Podolsky, and Rosen.  EPR was never mentioned
in any official academic setting. I immediately concluded that the
quantum-mechanical description of physical reality was incomplete, and
I made a note to think about completing it when I got tired of the
serious pursuits my teachers had set me on. (After my oral qualifying
exam, Roy Glauber advised me to stop spending so much time with Gordon
Baym. The senior members of my committee, Wendell Furry and Julian
Schwinger, seemed to agree.)

More than two decades later, in 1979, fifteen years after John Bell's
now-famous paper appeared, I learned about Bell's theorem through the
pages of the \emph{Scientific American}. I believe Tony Leggett had
tried to tell me about it a few years earlier, but I was too busy with
real physics to pay attention. I realized to my astonishment that the
more complete theory that EPR had convinced me would
someday be found to underly quantum mechanics, resolving all its
mysteries, either did not exist or, if it did, would be at least as
mysterious. In the three decades since then, I've devoted a
significant fraction of my intellectual efforts to pondering such
puzzles, mainly trying to boil them down to their simplest possible
forms.

\vskip 5pt \noindent{\bf Q2. What are the most pressing problems in the foundations of quantum
mechanics today?}

1. In the words of Chris Fuchs,\ft{I'm pleased to see that I cite the
   Master himself, in listing the most pressing of the problems.} ``quantum
   states: what the hell are they?'' Quantum states are not objective
   properties of the systems they describe, as mass is an objective
   property of a stone. Given a single stone, about which you know
   nothing, you can determine its mass to a high precision. Given a
   single photon, in a pure polarization state about which you know
   nothing, you can learn very little about what that polarization
   was. (I say ``was,'' and not ``is,'' because the effort to learn
   the polarization generally results in a new state, but that is not
   the point here.)

But I also find it implausible that (pure) quantum states are nothing
more than provisional guesses for what is likely to happen when the
system is appropriately probed. Surely they are constrained by known
features of the past history of the system to which the state has been
assigned, though I grant there is room for maneuver in deciding what
it means to ``know'' a ``feature.''\ft{This is an anti-QBist sentiment with,
however, a tip of the hat to QBism at the end of the sentence.}

Consistent historians (see also my answer to Question~16) maintain
that the quantum state of a system
\emph{is} a real property of that system, though its reality is with
respect to an appropriate ``framework'' of projectors that includes
the projector on that state. Since the reality of most other physical
properties is also only with respect to suitable frameworks, for
consistent historians the quantum state of a system is on a similar
conceptual footing to most of its other physical properties.\ft{ From a
Qbist perspective this is a deficiency.} Quantum
cosmologists maintain that the entire universe has an objective pure
quantum state.  I do not share this view. Indeed, I do not believe it
has a quantum state in any sense, since there is nothing (nobody)
outside the entire universe to make that state assignment.\ft{This is a
strictly QBist view, very concisely put.}  Well, I suppose it could be God, but 
why would he want to make state assignments? Einstein has assured us that he
doesn't place bets.\ft{A direct reference to the Dutch-book view of probability
dear to  QBists.}   (See also my answer to Question~4.)

2.  How clearly and convincingly to exorcise nonlocality from the
    foundations of physics in spite of the violations of Bell
    inequalities. Nonlocality has been egregiously
    oversold.\ft{Another position of QBists  and many others.}  On the
    other hand, those who briskly dismiss it as a naive error are
    evading a direct confrontation with one of the central
    peculiarities of quantum physics. I would put the issue like this:
    what can one legitimately require of an \emph{explanation} of
    correlations between the outcomes of independently selected tests
    performed on systems that no longer interact? (See also my answer
    to Question~8.)

3.  Is the experience of personal consciousness beyond the reach of
    physical theory as a matter of principle? Is the scope of physics
    limited to constructing ``relations between the manifold aspects
    of our experience,'' as Bohr maintained? While I believe that the
    answer to both question is yes, I list them as problems, because
    most physicists vehemently reject such views, and I am unable to
    explain to them why they are wrong in a way that satisfies me, let
    alone them.\ft{I would call this QBism.  My new QBist explanation of
    why most physicists are wrong comes closer to satisfying me, but
    most of them are as dissatisfied as ever.}

I regard this last issue as a problem in the interpretation of quantum
mechanics, even though I do not believe that consciousness (as a
physical phenomenon) collapses (as a physical process) the wave packet
(as an objective physical entity). But because I do believe that
physics is a tool to help us find powerful and concise expressions of
correlations among features of our experience, it makes no sense to
apply quantum mechanics (or any other form of physics) to our very
awareness of that experience.\ft{QBism.  Today I would sharpen it to ``It 
makes no sense for anybody to apply quantum mechanics to his or her 
very own awareness of that experience."  You can apply it to what you believe about
the experience of somebody else.}  Adherents of the many-worlds
interpretation make this mistake. So do those who believe that
conscious awareness can ultimately be reduced to physics, unless they
believe that the reduction will be to a novel form of physics that
transcends our current understanding, in which case, as Rudolf Peierls
remarked, whether such an explanation should count as ``physical'' is
just a matter of terminology.

I am also intrigued by the view of Schr\"odinger (in \emph{Nature and
the Greeks}) that it was a mistake dating back to the birth of science
to exclude us, the perceiving subjects, from our understanding of the
external world. This does not mean that our perceptions must be parts
of the world external to us, but that that those perceptions underlie
everything we can know about that world. (See also my answer to
Question~14.) Until the arrival of quantum mechanics, physics made
good sense in spite of this historic exclusion. Quantum mechanics has
(or should have) forced us to rethink the importance of the relation
between subject and object.\ft{I'm pleased and surprised to discover
myself writing this very QBist paragraph back in my dark pre-QBist days.}

\vskip 5pt \noindent{\bf Q3. What interpretive program can make the best sense of quantum 
mechanics, and why?}   

 My sympathies are with those, going all the way back to Heisenberg
 and Peierls, who maintain that quantum mechanics is a set of rules
 for organizing our knowledge with a view to improving our ability to
 anticipate subsequently acquired knowledge. By ``our knowledge,'' I
 mean my own knowledge combined with whatever other people are able to
 communicate to me of their own knowledge. I take this commonality of
 scientific knowledge to be one of the reasons why Bohr placed such
 emphasis on what can be expressed in ordinary language.\ft{Replace
 ``knowledge'' with ``belief'' and this becomes a very QBistic paragraph.}

To John Bell's ``Knowledge about what?'' I would say knowledge about
our percep\-tions---ultimately our direct, irreducible mental
perceptions, which can, of course, be refined by the use of
instruments devised for that purpose. To his ``Whose knowledge?'' I
would say knowledge of whoever is making use of quantum
mechanics. Different users with different perceptions may well assign
different quantum states to the same physical system. What consistency
requirements, if any, can be imposed on such descriptions, is an
entertaining question.\ft{QBism again, if you replace ``perceptions''
by ``experience'' and ``knowledge by'' ``belief''.}  I have had some
disagreements with some of my friends about this,\ft{Actually I
elicited a ferocious QBist attack, which I flatter myself had an
impact on the subsequent development of QBism.  See Caves, Fuchs, and
Schack, ``Conditions for compatibility of quantum state assignments'',
quant-ph/0206110, published as Phys. Rev. A66, 062111 (2002). I hereby
make a note to reexamine what, if anything, of my original point
survives my conversion.}  as described in ``Compatibility of state
assignments,'' which I cite here because it cannot be found in the
primary repository, arXiv, but only in the
\emph{Journal of Mathematical Physics} ({\bf 43}, 4560-66 (2002)).

My answer to ``Why?'' has to be inferred from my answers to most of
the other sixteen questions.

\vskip 5pt \noindent{\bf Q4. What are quantum states?}

 The first of my answers to Question~2 primarily says what quantum
 states are not. It is harder to say what they are. I am intrigued by
 the fact that if quantum mechanics applied only to digital quantum\
 computers, then the answer would be entirely straightforward. Quantum
 states are mathematical symbols. The symbols enable us to calculate,
 from the (explicit, unproblematic) prior history of a collection of
 Qbits---I commend to the reader this attractive abbreviation of
 ``qubit''---the probabilities of the readings ($0$ or $1$) of a
 collection of one-Qbit measurement gates to which the Qbits are then
 subjected. This procedure is made unambiguous by the rule that a Qbit
 emerging from a one-Qbit measurement gate reading $0$ (or $1$) is
 assigned the state $|0\rangle$ (or $|1\rangle$). This makes it
 possible to assign initial states with the help of one-Qbit
 measurement gates. Additional rules associate specific unitary
 transformations of the states of the Qbit(s) with the action of the
 other subsequent gates that appear in a computation.

Quantum states, in other words, are bookkeeping tools that enable one
to calculate, from a knowledge of the initial preparation and the
fields acting on a system, the probability of the outcomes of
measurements on that system.\ft{My QBist friends don't like this.  At a minimum
they would replace ``knowledge of" with ``belief about".} This is what I take to be the Copenhagen
interpretation of quantum mechanics. (I hereby renounce my earlier
summary of Copenhagen, widely misattributed to Richard Feynman, as
``shut up and calculate.'') If the only application of quantum
mechanics were to the operation of digital quantum computers, there
would be no ambiguity or controversy about Copenhagen.

The Copenhagen view fits quantum computation so well that I am
persuaded that quantum states are, even in broader physical contexts,
calculational tools, invented and used by physicists to enable them to
predict correlations among their perceptions.\ft{My QBist friends
would not object to this, particularly if ``perceptions" was changed to
``experience".} I realize that others have used their experience with
quantum computation to make similar arguments on behalf of many worlds
(David Deutsch) and consistent histories (Bob Griffiths). I would
challenge them to make their preferred points of view the basis for a
quick practical pedagogical approach to quantum computation for
computer scientists who know no physics, as I have done with
Copenhagen in my quantum-computation book. The approach to quantum
mechanics via consistent histories in Griffiths's book, while
something of a tour de force, does not strike me as either quick or
practical.

\vskip 5pt \noindent{\bf Q5. Does quantum mechanics imply irreducible 
randomness in nature?}

Yes. But ``in nature'' requires expansion. A more precise formulation
would be that quantum mechanics implies irreducible randomness in the
answers to most of the questions that we can put to nature. The
probability of a photon that has emerged from a vertically oriented
sheet of polaroid getting through one oriented at forty-five degrees
from the vertical is irreducibly one half, as is the probability of a
slow-moving mu meson turning into an electron and a pair of neutrinos
in the next microsecond-and-a-half. ``Irreducible'' means there is
nothing we can condition the probabilities on that would sharpen them
up.\ft{I seem here to be flirting with objective probabilities.}

Can you exploit quantum physics to make an ideal random-number
generator?\ft{The QBist answer to this question is not obvious.
It is not even clear what, if anything, my question means to
one who takes a subjective view of probability.}  A distinguished
Cornell computer scientist once made the long trek from the
Engineering Quad to my physics-department office in the heart of the
Arts College to ask me this question. He had been told this by a
student, and didn't believe him. I said the student was right. I don't
think he believed me either.

\vskip 5pt \noindent{\bf Q6. Quantum probabilities: subjective or 
objective?}\ft{My answer is an interesting (to me) pre-Stellenbosch
attempt to reconcile my IIQM with Fuchs's early QBism.}

 In a message in a bottle that I tossed into the sea about fifteen
years ago---the ``Ithaca Interpretation of Quantum Mechanics''
(IIQM)---I firmly declared quantum probabilities to be objective
properties of the physical world. The bottle was noticed by Chris
Fuchs, who introduced me to subjective probabilities and to his
collaborators Carl Caves and R\"udiger Schack. I found their point of
view so intriguing that I have left the bottle adrift ever since, but
in thinking about it today, I wonder why I was so readily persuaded
that their view of probability was incompatible with mine.

In declaring quantum probabilities to be objective, I had in mind two
things. First, that the role of probability in quantum mechanics is
fundamental and irreducible. Probability is not there just as a way of
coping with our ignorance of the underlying details, as in classical
statistical mechanics. It is an inherent part of how we can understand
and deal with the world. Second, that probabilistic assertions are
meaningful for individual systems, and not just, as many physicists
would maintain, for ensembles of ``identically prepared'' systems. I
believe Fuchs {\it et al.\/} would agree with both propositions.

I also explicitly rejected Karl Popper's promotion of ``propensities''
into objective properties of the systems they describe. It was not my
intent to reify probability, or if it was---fifteen years later it is
hard to be sure---I hereby disassociate myself from the foolish person
I might then have been. Admittedly, my IIQM motto that
``correlations have physical reality'' (though correlata do not)
sounds dangerously like a Popperian reification of probability. But it
is not. In my two IIQM papers, I used the phrase ``has
physical reality'' to mean ``can be accounted for in a physical
theory,'' particularly when I insisted that conscious experience has
reality, but not physical reality.

Thinking about this today, I see that to be compatible with the point
of view of Fuchs {\it et al.}, I should also have maintained that
correlations have physical reality but not reality. ``Physical
reality'' is not, as I seem to have implicitly maintained fifteen
years ago, just a subset of ``reality.'' Neither is contained in the
other. Conscious awareness belongs to reality and not to physical
reality, but correlation belongs to physical reality and not to
reality. Putting it like that, I now see that this goes a way toward
reconciling the IIQM not only with Fuchs {\it et
al.}, but also with Adan Cabello's demonstration that whatever the
sense in which correlations have physical reality, it cannot be that
their values are EPR ``elements of reality.''

So I would say that quantum probabilities are objective in the sense
that they are unavoidable. They are intrinsic features of the quantum
formalism---not just an expression of our ignorance. And they apply to
individual systems and are not just bookkeeping devices for
cataloguing the behavior of ensembles of identically prepared systems.

But because quantum mechanics is our best strategy for organizing our
perceptions of the world, quantum probabilities have a strategic
aspect. Strategy implies a strategist, and in that sense quantum
probabilities are subjective.\ft{I would like to think this is a QBist 
sentiment.}

Strategic as the use of probability may be, the fact that a free
neutron has a slightly less than fifty-fifty chance of decaying within
the next ten minutes strikes me as just as objective a property of the
neutron as the fact that its mass is a little less than 1,839 times
the mass of an electron. Of course, one can, and some of my friends
do, conclude from this that dynamics itself (in which mass is a
parameter, and out of which emerges the half-life) is as subjective a
matter as probability. Wary as I am of reification, I'm not ready to
take that step.\ft{A rare anti-QBist position.}

\vskip 5pt \noindent{\bf Q7. The quantum measurement problem: serious 
roadblock or dissolvable pseudo-issue?}

 It's a pseudoissue. But I have not dissolved it entirely to my
 satisfaction. So while I see no roadblock, I do feel the need to
 drive slowly past some unfinished construction, attending to signals
 from the people with flags.

Today ``the quantum measurement problem'' has almost as many meanings
as ``the Copenhagen interpretation.'' I mention only two of them. The
first is how to account for an objective physical process called the
collapse of the wave function, which supersedes the normal unitary
time evolution of the quantum state in special physical processes
known as measurements. I believe that this version of the problem is
based on an inappropriate reification of the quantum state. So are
efforts to eliminate the special role of measurement through dynamical
modifications in standard quantum mechanics that make an appropriate
rate of collapse an ongoing physical process under all conditions.

The quantum state is a calculational device, enabling us to compute
the probabilities of our subsequent experience on the basis of earlier
experiences.  Collapse is nothing more than the updating of that
calculational device on the basis of additional
experience.\ft{Quintessential QBism!} This point of view is the key to
resolving this form of the quantum measurement problem. I look forward
to the day when some clear-headed gifted writer has spelled it out so
lucidly that everybody is completely convinced that there is no such
problem. (I'm convinced. But I'm not completely convinced.\ft{Now I am.})

A second question going under the name ``quantum measurement problem''
is whether there can be quantum interference between quantum states
that describe macroscopically distinct physical conditions (sometimes
called ``cat states''). If such interference is not just hard to
observe but strictly absent, then quantum mechanics must break down in
its answers to questions of sufficient complexity, asked of systems of
sufficient size. Size alone is not the issue, since quantum mechanics
works brilliantly in accounting for all kinds of classically
inexplicable behavior in the gross behavior of bulk materials. Indeed,
the appropriate definition of ``macroscopic'' in this setting is far
from obvious.

The fact that the unavoidable entanglement of a macroscopic system
with its environment renders manifestations of quantum interference
effectively unobservable is a good practical rejoinder to those who
seek an answer from a macroscopic breakdown of quantum mechanics. But
decoherence does not directly address the question of whether anything
actually changes when the superposition is replaced by a mixed state,
beyond an abstract representation of our practical ability to acquire
knowledge. And it is subject to the same kinds of time-reversal
problems that plague statistical-mechanical derivations of the second
law.

Seeing quantum interference effects with carbon-60 molecules is an
experimental tour de force. But I would have been astonished if
interference had been demonstrably absent. My impression is that those
who did the experiment did not expect it to reveal a breakdown of
quantum mechanics. They did it because it was there, like Mount
Everest, challenging somebody to take it on.

\vskip 5pt \noindent{\bf Q8. What do the experimentally observed violations of
 Bell's inequalities tell us about nature?}

They tell us something strange about correlations in the outcomes of
certain sets of local tests, independently chosen to be performed on
far-apart noninteracting physical systems, which may have interacted
in the past but no longer do. Prior to Bell's analysis of such
quantum-theoretic correlations (and the experimental confirmation of
those theoretic predictions), it seemed reasonable to assume that
correlations in the outcomes of such tests could find an explanation
in correlations in the conditions prevailing at the sites of the
tests. Such local conditions can include individual features of the
locally tested system, acquired at the time of its past interaction
with the other systems; the conditions can also include the weather at
the place of the test, the time of each local test, and so on.

Such local explanations can indeed be constructed for any single
choice of which local test to perform on each system. But if there is
more than one choice of test for each system, then there can be
circumstances (revealed by a violation of an appropriate Bell
inequality) in which no single explanation, based on correlation in
the locally prevailing conditions, works for all possible choices of
local test, even if the choices of local test are made randomly and
independently in each local region. This is strange, because the local
conditions prevailing at the site of any particular test cannot depend
on a random choice of what test to perform far away from that site.

Failure of a Bell inequality fatally undermines the view that all the
correlations in all the possible tests can find a single explanation
in terms of correlations in conditions at the sites of the tests. The
conclusions people draw from this vary widely. Those who conclude that
the choice of what test to perform in one region does affect the
prevailing conditions in the other regions (as it does explicitly in
the de~Broglie--Bohm pilot-wave interpretation) have embraced
nonlocality.

A more conservative conclusion is that it is unreasonable to demand a
single explanation that works not only for the choices of test that
were actually made in each region, but also for the choices of test
that might have been made but were not. This is the conclusion of that
subset of the quantum-information community with which I
sympathize. It is also the conclusion of consistent historians (see my
answer to Question~16), but their apparent conservatism hides their
ontologically radical insistence that all the explanations give
correct accounts of the tests to which they apply, subject to the
proviso that you cannot combine ingredients of one explanation with
those of any other, since their validity is in general relative to
different ``frameworks.''

I like Asher Peres's conclusion that unperformed tests have no
outcomes: it is wrong to try to account for the outcomes of all the
tests you might have performed but didn't. This too is more radical
than it appears, since recent versions of Bell's theorem (inspired by
Danny Greenberger, Mike Horne, and Anton Zeilinger) show that the
outcome of the test you actually performed is incompatible with each
and every possible set of outcomes for all the tests you might have
performed but didn't. This adds a word to Asher's famous title:
``Unperformed experiments have no \emph{conceivable} results.''

That addition makes his point just a little harder to swallow. But
swallowing becomes easier again if I expand Asher's title further to
``Many different sets of unperformed experiments have no conceivable
sets of results, \emph{if} the result for each local test has to be
exactly the same in every set of results in which that particular
local test appears.'' (The expanded title itself, however, is harder
to swallow.) What can it mean to impose such consistency on sets of
conceivable data associated with different choices of sets of local
tests, when only one set of tests was actually performed?

So for me, nonlocality is too unsubtle a conclusion to draw from the
violation of Bell inequalities. My preference is for conclusions that
focus on the impropriety of seeking explanations for what might have
happened but didn't. Evolution has hard-wired us to demand such
explanations, since it was crucial for our ancestors to anticipate the
consequences of all possible contingencies in their (classical)
struggles.

(See also the second of my answers to Question~2.)

\vskip 5pt \noindent{\bf Q9. What contributions to the foundations of
quantum mechanics have or may come from quantum-information theory?
What notion of `information' could serve as a rigorous basis for
progress in foundations?}

I agree with Heisenberg and Peierls that the quantum formalism is a
tool we have discovered to express the information we have acquired
and the consequences of that information for the content of our
subsequent acquisition of information. To the extent that it sharpens
and systematizes this point of view, I believe that quantum
information theory is the most promising and fruitful foundational
approach.\ft{Replace ``information'' by ``belief'' and this is QBism.}

Beyond this, applying the quantum formalism directly to the processing
of information itself may get us closer to the heart of what quantum
mechanics is all about, than can the informationally less subtle
problems addressed in more traditional physical applications of
quantum mechanics. At the very least, it provides a refreshingly
different set of examples of quantum phenomena.

I am not expert enough in quantum (or classical) information theory to
have an opinion on the definition of information most likely to shed
light on foundational questions. Slogans like ``It from bit'' are fun,
but don't tell me much without considerable (yet to be provided)
expansion. It seems to me that any foundationally illuminating concept
of information must be explicit about both the possessors of the
information and the content of that information.\ft{QBism is explicit: The possessor of
the belief is the agent using quantum mechanics to assign
probabilities; the content of the belief comes from the experience of
the agent.} As John Bell put it, ``Whose information?'' and
``Information about what?''

(See also my answer to Question~16.) 

\vskip 5pt \noindent{\bf Q10. How can the foundations of quantum mechanics benefit from 
approaches that reconstruct quantum mechanics from fundamental 
principles? Can reconstruction reduce the need for interpretation?}

It is wonderful that all of special relativity follows from the
principle that no physical behavior can distinguish among frames of
reference in different states of uniform motion, combined with the
realization that the simultaneity of events in different places is a
convention that can differ from one frame of reference to another. Can
the rest of physics---in particular quantum mechanics---be reduced to
so economical a set of assumptions?

I doubt it. Even the foundations of special relativity are not
captured as compactly as I just claimed. I failed, for example, to
mention the assumptions of spatial and temporal homogeneity, and of
spatial isotropy. And the fundamental notion of an ``event''---a
phenomenon whose spatial and temporal extent we can ignore for
purposes of the topic currently under discussion---might strike some
as irritatingly vague, bringing ``us'' into the story in a way physics
traditionally (and, I increasingly believe, wrongly\ft{Definitely a QBist
sentiment.}) tries to avoid. And just what are these human artifacts
called ``clocks'' that play so fundamental a role in the story? In
short, it's not as simple as advertised.

Yet quantum mechanics does seem to be floating in the air, in a way
that makes relativity seem quite anchored. At least the basic
conceptual ingredients of relativity have at first glance a direct
intuitive correspondence with familiar phenomena in our immediate
experience. The complicating issues for relativity emerge only when
one insists on sharpening up these intuitions. In contrast, the basic
ingredients of quantum mechanics---states, superpositions, and their
linear evolution in time---bear not even a vague relation to anything
in our direct experience, while measurement---the only thing that ties
the subject to the ground---seems to introduce what John Bell derided
as ``piddling laboratory operations'' at too fundamental a level.

I'm glad people are attempting to reconstruct quantum mechanics from
(a few) fundamental principles, but I'm skeptical that they'll succeed
without slipping into at least one of their principles something just
as much in need of interpretation. The reason I'm nevertheless glad is
that having a new and strikingly different formulation of the really
puzzling stuff can sometimes be a useful step toward untangling the
puzzle

\vskip 5pt \noindent{\bf Q11. If you could choose one experiment, 
regardless of its current technical feasibility, to help answer a
foundational question, which one would it be?}

The foundational issues about quantum mechanics that perplex me are
all predicated on the assumption that the theory is correct. I would
like to be able to make better sense of what it says. I am not
persuaded that my perplexity is so acute that I should seek the answer
in a breakdown of the theory. Therefore I would not expect any
experimental test to shed light on a foundational question.

I exclude here the possibility that a crucial foundational issue might
be associated with an application of the theory so intricate that the
relevant calculation might be too difficult to perform, thereby
requiring an experimental test. It does seem to me that all the
puzzling features of the theory emerge full-blown in its most
calculationally elementary applications.

This is not to say that the breakdowns of quantum mechanics suggested
by some interpretations are not worth exploring through
experiment. (See also my answer to Question~7.) I expect quantum
mechanics to break down at some scale. Indeed, I find it amazing, in
view of the body of data that gave rise to it, that it seems to be
working perfectly well within the atomic nucleus and even within the
nucleon. This lends support to viewing quantum mechanics as a ``mode
of thought,'' as Chris Fuchs and R\"udiger Schack once put it, rather
than as a description of the world.\ft{A tip of the hat to the two leading QBists .}

So I would be surprised (and rather disappointed) if foundational
issues were settled by observing a breakdown of quantum mechanics. I
would expect them to be settled by our acquiring a deeper
understanding of the existing theory, within its domain of validity.

\vskip 5pt \noindent{\bf Q12. If you have a preferred interpretation 
of quantum mechanics, what would it take to make you switch sides?}

My intuitions about the nature of quantum mechanics are not coherent
enough to add up to anything I would dignify with the term
``interpretation.''  Admittedly, shortly after turning sixty, I did
write a few papers setting out what I called the Ithaca Interpretation
(see also my answer to Question~6).  But I was young then, innocent,
and overly willing to sacrifice an accurate phrase for an entertaining
one.

One of those papers made its argument under the banner of Bohr's
statement that the purpose of our description of nature is ``only to
track down, so far as it is possible, relations between the manifold
aspects of our experience.'' When I wrote the paper, the crucial word
for me was \emph{relations}. My motto was \emph{correlations without
correlata}. What led me to stop giving physics colloquia on the IIQM
after only a year was the obvious question: ``Correlations between
what?'' Abner Shimony aptly complained that the Ithaca Interpretation
``had no foreign policy.''  Exchanges with Chris Fuchs persuaded me
that just as important as \emph{relations} was \emph{our experience},
which I was too ready to hide beneath the same rug under which I had
(correctly) swept the problem of consciousness.

So insofar as I had a preferred interpretation in 1998, what persuaded
me that it was, at best, insufficiently developed was somebody making
me aware of some interesting ideas that hadn't occurred to me. It
remains entirely possible that some wise, imaginative, and readable
person may in the future lure me away from the position I am trying to
sketch in my answers to these questions.\ft{That position is so close
to QBism that I'm not sure that in Stellenbosch my QBist friends lured me very far.}

To make me switch to some interpretations I now reject would require a
breakdown of quantum mechanics along lines suggested by these
currently unpalatable points of view. Of course, at that point they
would no longer be interpretations of existing theory, but alternative
theories. To convert me to Bohmian mechanics, for example, I would
have to see clear evidence of particles that were not in ``quantum
equilibrium.'' Without that breakdown of orthodox quantum mechanics,
the reintroduction of particle trajectories seems an unnecessary
complication that raises questions at least as vexing as those raised
by the orthodox theory. To convert me to the view that ``wave-function
collapse'' was a real physical process and not just an updating of
expectations on the basis of new information, I would have to see
convincing evidence of deviations from quantum probabilities produced
by Ghirardi--Rimini--Weber--Pearle ``hits.''

A simple nontrivial example of a history containing many different
times that exactly satisfied the consistency conditions might persuade
me to take another look at consistent histories (see my answer to
Question~16).

\vfil\eject
\vskip 5pt \noindent{\bf Q13. How do personal beliefs and values
influence one's choice of interpretation?}

The belief that physics is, or ought to be, the whole story surely
plays a role. Those who believe that physics describes the external
world \emph{as it relates to us} have an interpretive flexibility
unavailable to those who insist that ``we'' have no place in the story
except as complex physical systems.\ft{QBists!} (See also the third of my answers
to Question~2.)

I have the impression that those physicists who believe in God tend,
perhaps unsurprisingly, to take a more strongly realistic view of the
abstractions that make up the quantum formalism than do many of us who
take an atheistic view of the world.

There are also those who maintain that while God does not exist,
Physical Law does. Since I agree with the first half of this
proposition, I would not call them idolatrous. But others might.

Values (as opposed to beliefs) are harder to identify. I sometimes
detect them in the attitudes of those who believe in, or search for,
hidden-variables models of quantum mechanics. I have heard ringing
declarations about the nature of science, exhortations not to give up
the good fight, and expressions of scorn for contemporary
obscurity. (See also my answer to Question~15.)

\vskip 5pt \noindent{\bf Q14. What is the role of philosophy in advancing our understanding of 
the foundations of quantum mechanics?}

If quantum mechanics is correct, or even if it is only correct to a
high degree of accuracy in some yet-to-be-delimited domain, then
everything in quantum foundations counts as philosophy. Let me
rephrase the question: what role have professional philosophers played
in advancing our understanding of the foundations of quantum
mechanics? I do not count as ``philosophers'' professional
philosophers who are also professional physicists, and I count as
``professional'' anybody with a Ph.D.\ in the field.

When I got into this business thirty years ago, I had hoped that
philosophers would bring to the conversation their historical
expertise in the Big Questions. What is the nature of human knowledge?
How do people construct a model of the world external to themselves?
How does our mental organization limit our ability to picture
phenomena? How does our need to communicate with each other constrain
the kinds of science we can develop? Those kinds of questions.

To my disappointment, it seems to me that professional philosophers
prefer to behave as amateur physicists. They don't try to view the
formalism as part of a Bigger Picture. On the contrary, they seem to
prefer to interpret it more literally and less imaginatively than many
professional physicists. Because they are less proficient than
physicists in using the tools of physics, they tend not to do as good
a job on these narrower matters. They often come through as naive and
unsophisticated.

So I would say that up to now, professional philosophers have not
played a significant role in advancing our understanding of quantum
foundations. I would not (and could not) discourage them from working
in quantum foundations. But I would urge them to keep their eyes on
the Big Questions.  (See also my answer to Question~15.)

\vskip 5pt \noindent{\bf Q15. What new input and perspectives for the 
foundations of quantum mechanics may come from the interplay between
quantum theory and gravity/relativity, and from the search for a
unified theory?}

My guess is that an understanding of the connection between gravity
and quantum mechanics will have to await new input and perspectives
from the foundations of both disciplines. Space and time in quantum
field theory are classical parameters. They're on \emph{our} side of
the subject--object boundary.\ft{The distinction between subject
(agent) and object (world external to the agent) is a fundamental part
of QBism.  Schr\"odinger also makes much of it, particularly (but not
exclusively) in {\it Nature and the Greeks.}  See also my {\it
Commentary} in the July 2012 {\it Physics Today} and my remarks in the
December 2012 issue about the letters it elicited.}  Extrapolating
them down to sub-nucleon levels---let alone to the Planck
scale---strikes me as unwarranted and even arrogant. (I note with
interest a hint of some personal values here. Compare my answer to
Question~13.)  Spatial and temporal coordinates describe the readings
of our instruments.

I'm just as skeptical about quantum cosmologists applying quantum
mechanics to the universe as a whole. For quantum mechanics to make
sense, there has to be an inside (``the system'') and an outside
(``us'').\ft{More QBism.}

So insofar as gravity is a theory of the structure of space-time, I'd
be surprised if real progress were made in incorporating it into
quantum theory without a more thoughtful and (dare I say it?)
\emph{philosophical} examination of the foundations of both fields.

\vskip 5pt \noindent{\bf Q16. Where would you put your money when it 
comes to predicting the next major development in the foundations of
quantum mechanics?}

Let me put the question in a more manageable form: what was the last
major development in the foundations of quantum mechanics? (It remains
basically the same question, since none of the developments that
follow have been broadly accepted as the most illuminating way to look
at the subject.)

I would nominate for the most important recent development the
application of quantum mechanics to the processing of information,
starting with the invention of quantum cryptography by Bennett and
Brassard in 1984, continuing with the development of quantum
computation, and the fascinating efforts of Chris Fuchs to make a
coherent whole out of it all.\ft{QBism is at the top of the list.} As
runner-up, I would cite the study of pre- and postselected ensembles
by Aharanov and his collaborators, and (perhaps---I still lack a good
feeling for it) the ensuing notion of weak measurement. In third
place, I would put the consistent-histories point of view, as put
forth by Bob Griffiths.

What all three of these developments have in common is that they are
standard quantum mechanics applied in highly nonstandard settings. In
this respect, they are all conservative approaches to quantum
foundations. They use the orthodox theory to answer simple questions
that it had never before occurred to anybody to ask. The answers
provide intriguing new perspectives on the theory.

Because the last of the three seems to have been widely ignored in the
quantum-foundations community and is unrepresented among the authors
of this volume, I'll say a little about it. (My old friend Pierre
Hohenberg has tried valiantly to get me to take this stuff
seriously.\ft{He disapproves of my recent interest in QBism.}
Pierre and I were in both college and graduate school together, but in
all those years nobody ever warned me to stay away from him; see my
answer to Question~1. Maybe somebody should have.)

Consistent historians offer an unusual fusion of collapse and
no-collapse points of view.  Underlying their weltanschauung is an old
formula of Aharanov, Bergmann, and Lebowitz (ABL), which compactly
gives the probabilities of the outcomes of a whole sequence of (von
Neumann) measurements carried out at different times on a system in a
given initial state.\ft{Bob Griffiths tells me that the formula was
published by Wigner a year before ABL.}  Prior to its
reinterpretation\ft{Griffiths also made me realize that I should have
said ``drastic reinterpretation''. I believe that I did in an earlier
draft, since I was surprised to discover, when he complained, that
``drastic'' was not in the book.}  by consistent historians, the {ABL}
formula was understood to be an expression of the fact that
immediately after any particular measurement, the state of the system
collapses according to the standard Born rule; this postcollapse state
then evolves under the unitary dynamics until the next measurement in
the sequence produces another collapse. Unitary evolution, followed by
measurement and collapse, followed by more unitary evolution, followed
by more measurement and collapse, and so on.

Consistent historians eliminate measurement and collapse from the
story by reinterpreting these probabilities to be probabilities of
what I would call \emph{actual states of being}---called
histories. These histories (or, more accurately, the subset of them
deemed ``consistent,'' as noted below) have nothing to do with
measurement outcomes. For consistent historians the {ABL}
formula is thus more fundamental and broader in scope than the Born
rule. The Born rule can be extracted from the consistent historians'
version of {ABL} in some very special cases, but
measurement vanishes from {ABL} in the general case, which
according to consistent historians gives probabilities not of
measurement outcomes but of actual states of being.

How can they get away with this vast extension of actuality to
entities whose nonexistence lies at the very heart of conventional
quantum mechanics? Easily! They do it by forbidding the extension
whenever it gets you into trouble; they impose stringent consistency
conditions on the probabilities appearing in any candidate for a valid
history. Any history that meets these consistency conditions can
describe the probabilities of an actual state of being, and not the
mere outcomes of a set of piddling laboratory operations. Any history
that violates the consistency conditions is utter nonsense---not a
history at all, and certainly not a description of actual states of
being.

As one might expect, there can be many distinct histories, all of
which meet their own internal consistency conditions, although the
state of being that combines the actual states of being associated
with more than one of those histories need not satisfy its own
internal consistency conditions. When this happens, the combination of
the two actual states of being is not an actual state of being.

Rather than concluding from this that the project is dead in the
water, the consistent historians elevate it to a fundamental
ontological principle. Reality is multifaceted. There can be this
reality or there can be that reality, and provided you refrain from
combining actualities from mutually inconsistent realities, all of the
incompatible realities have an equally valid claim to actuality. This
tangle of mutually incompatible candidates for actuality (associated
with different ``frameworks'') constitutes the no-collapse side of
consistent histories. The collapse side lies in the fact that each of
these peacefully coexisting mutually exclusive actualities is
associated with what from the orthodox point of view (which consistent
historians reject) would be a sequence of measurements and Born-rule
collapses.

This multiplicity of incompatible realities reminds me of special
relativity, where there is time in this frame of reference and time in
that frame of reference, and provided only that you do not combine
temporal statements valid in two different frames of reference, one
set of temporal statements is as valid a description of reality as the
other.

But I am disconcerted by the reluctance of some consistent historians
to acknowledge the utterly radical nature of what they are
proposing. The relativity of time was a pretty big pill to swallow,
but the relativity of reality itself is to the relativity of time as
an elephant is to a gnat. (Murray Gell Mann, in his talk of ``demon
worlds,'' comes close to acknowledging this, yet he dismisses much
less extravagant examples of quantum mysteries as so much
``flapdoodle.'')

\vskip 5pt \noindent{\bf Q17. What single question about the
foundations of quantum mechanics would you put to an omniscient
being?}

I'd ask, ``How has the uncertainty principle altered the `omni' of
your omniscience?''

Joking aside---but it wasn't really a joke---I have trouble imagining
an omniscient being. Let me rephrase the question: if you could be
frozen for 150 years and revived intact, what question would you ask
physicists when you woke up?

I'd ask something like this: 

Is the fundamental physics of a system still described in terms of
quantum states that evolve linearly in time and that specify
probabilities of the outcomes of tests that we can perform on that
system? If so, is anybody puzzled by the meaning of this conceptual
structure? If not, is there general agreement on the meaning of the
structure that replaced it?

In early twenty-first-century terms: has the structure of quantum
mechanics survived intact for a century and a half? If so, are there
still foundational problems? If not, are there still foundational
problems?

I chose 150 years because a century might not be long enough to get an
interesting answer. But I also worry that physicists two centuries
from now, no matter how I phrased the question, might not understand
it. It might elicit only polite bewilderment, just as a pressing
aether-theoretic query at the end of the nineteenth century might seem
not only irrelevant but downright incomprehensible to a physicist of
the early twenty-first.

There are two possible grounds for future bewilderment at my
question. One is that quantum mechanics will have been discovered, as
Einstein always hoped, to be a phenomenology based on a more
fundamental view of the world, which is more detailed and more
intuitively accessible. This strikes me as unlikely, because John Bell
showed that any theory detailed enough to satisfy certain common-sense
yearnings would also have to contain instantaneous
action-at-a-distance. (See my answer to Question~8.) So while the
discovery of a more fundamental view of the world during the next
century and a half seems entirely possible, I'd be surprised if the
new theory turned out to be more intuitive than our current
understanding.

An appropriate timescale for the survival of quantum mechanics is set
by the fact that its basic conceptual machinery has suffered no
alterations whatever, beyond a little tidying-up, for over eighty
years. Not a bad run when you compare what happened to fundamental
knowledge between 1860 and 1940, though not close to the more than two
centuries that classical mechanics remained the fundamental theory. So
the persistence of the same basic formalism for another 150 years
seems at least plausible.

Even so, my question might elicit mid-twenty-second-century
bewilderment, because after several more generations of physicists,
chemists, biologists, engineers, and computer scientists had worked
with the theory, it might finally, in Feynman's words, have become
obvious to everybody that there's no real problem. We early
twenty-first-century people, who believed there ought to be a better
way to understand the theory, will then have been consigned to the
same dustbin of history as the early twentieth-century aether
theorists.

I hope that's not how it works out. It is, for example, now possible
to articulate the nature of the wrong thinking that made relativity
seem shockingly counterintuitive to many people during its early
years. People had simply deluded themselves into believing that there
was something called ``time'' that clocks recorded, rather than
recognizing that ``time'' was a remarkably convenient abstraction---I
would say an ingenious abstraction, except that nobody set out
deliberately to invent it---that enables us to talk efficiently and
even-handedly about the correlation among many different kinds of
clocks.

There is now no generally agreed-upon key to dissolving the puzzlement
that quantum mechanics engenders today in many of us. (For that
matter, I have encountered otherwise sensible physicists who disagree
with the above resolution of the puzzles of relativity.) I would hope
that within the next 150 years, such a key might be found that almost
everybody would agree clarifies the character of the theory, in
contrast to today's state of affairs, where no school of thought
commands more than ten percent of the population, except for those who
maintain---but can they really mean it?---that there is nothing to be
puzzled about.

\bye